\documentclass[fleqn]{mnras}

\usepackage[T1]{fontenc}
\usepackage{ae,aecompl}

\usepackage{graphicx}

\graphicspath{ {./figs/} }

\newcommand{\HI}{\textrm{H}\,\textsc{i}}
\newcommand{\Halpha}{\textrm{H}\,$\alpha$}
\newcommand{\LCDM}{$\Lambda$CDM}
\newcommand{\kms}{km\,s$^{-1}$}

\title[Morphological properties of LV galaxies]{Morphological properties of galaxies 
in different Local Volume environments}
\author[Karachentsev, Kaisina, Makarov]{
I.D. Karachentsev$^1$\thanks{ikar@sao.ru}, 
E.I. Kaisina$^1$
and D.I. Makarov$^1$
\\
$^1$ Special Astrophysical Observatory of the Russian Academy of Sciences, 
Nizhny Arkhyz, Karachai-Cherkessian Republic
} 

\date{Accepted XXX. Received YYY; in original form ZZZ}
\pubyear{2018}

\begin{document} 
\label{firstpage}
\pagerange{\pageref{firstpage}--\pageref{lastpage}}
\maketitle

\begin{abstract}
We consider an all-sky sample of 1029 Local Volume (LV) galaxies situated within a distance of 11~Mpc. 
Their majority have precise distances, estimates of hydrogen mass fraction and star-formation rate derived from far-ultraviolet or \Halpha{} fluxes. 
To describe an environment, we attribute two dimensionless values: 
the density contrast created by the most significant neighbour and the local density contrast produced by all neighbours within a separation of 1~Mpc. 
The hydrogen mass fraction exhibits a weak effect of \HI{} deficiency being the most pronounced for dwarf irregular galaxies.
The specific star-formation rate (sSFR) is more sensitive to the environment than the hydrogen mass fraction. 
Almost all (99 per cent) LV galaxies have their sSFR below $-9.4$ dex (yr$^{-1}$).
We notice that irregular dwarfs as well as late-type bulgeless galaxies are capable to reproduce their stellar mass with the observed sSFR over the cosmic time.
Thus, the transformation of gas into stars in dIrs and spiral disks is rather sluggish unlike that in E, S0, dSph galaxies, whose star-formation history has been stormy and short. 
Scatter of SFR(\Halpha{})-to-SFR(FUV) ratio increases from Sc, Sd, Sm galaxies towards BCD, Im, Ir types that favours the idea of bursty star formations in low-mass galaxies. 
However, this bursty activity is caused rather by internal processes than by an external tidal action. 
A fraction of quenched E, S0, dSph galaxies increases from $\sim5$ per cent in the field up to $\sim50$ per cent in the densest regions.
\end{abstract}

\begin{keywords}
galaxies: formation -- galaxies: dwarf -- galaxies: star formation
\end{keywords}

\section{Introduction} 

The \LCDM{} standard cosmological model effectively explains the origin and 
evolution of the large-scale structure of the Universe.
However, with transition to $\leq 1$-Mpc scales, predictions of the standard 
model begin to diverge from observation data.
One of the most well-known discrepancies is the observed shortage (dozens of 
times!) of a number of dwarf companions around normal-mass galaxies as compared 
with the results of numerical modelling in the framework of \LCDM{} (Moore 
et al. 1999, Klypin et al. 1999). 

To test numerical models, one needs a sample of galaxies limited by their 
distance rather than flux (apparent magnitude).
The difference between two kinds of galaxy selection is enormous, e.g., the 
famous Revised Shapley-Ames Catalog (Sandage \& Tammann 1981)
with 1246 brightest galaxies and a catalogue of 1000 nearest galaxies overlap 
by 8 per cent only.
According to Peebles (1993), compiling a representative sample of nearby 
galaxies located within a fixed distance is an important task for observational 
cosmology on small scales.

The global properties of galaxies, such as atomic hydrogen abundance or star-
formation rate, have been considered by many authors 
(Huang et al. 2012, Karachentsev \& Kaisina 2013, Knobel et al. 2015, Knapen et 
al. 2015, Beygu et al. 2016, Watkins et al. 2017) 
using samples of different composition and depth.
However, from our point of view, a representative sample of the nearest galaxies 
has an undeniable advantage over the others in this respect, since various 
effects of observational selection in it are easier to analyse and account for.

There is extensive literature on the dependence of integrated parameters of 
gaseous and stellar components of galaxies on the density of their environments.
In recent years, different scenarios of transformation of gas into stars in 
galaxy disks and bulges were discussed by 
Birrer et al. (2014), Schawinski et al. (2014), Pipino et al. (2015), 
Rodriguez-Puebla et al. (2017), Semenov et al. (2017), Barsanti et al. (2018).
According to many data, the process of morphological evolution of a galaxy 
essentially depends on its baryon mass which increases with time as galaxies 
occasionally merge.
However, it remains unclear what the role of diffuse intergalactic medium 
is in the global evolution of normal and dwarf galaxies.
In accordance with predictions of the standard cosmological model, the basic 
amount of cosmic baryons is distributed outside optical borders of galaxies 
as warm plasma, the large-scale structure and temperature of which is still 
the target of speculations.
Systematic analysis of observation data on galaxies in the reference sample 
of the Local Volume can facilitate a solution of the issue specified.

\section{The Local Volume sample}                                                                 
                                                                                                  
The first step to create a distance-limited sample of nearby galaxies was                         
made by Kraan-Korteweg \& Tammann (1979), who compiled a list of 179                              
galaxies with expected distances within 10~Mpc. Later, Karachentsev (1994)                        
and Karachentsev et al. (2004) expanded this list up to 226 and 450 galaxies,                     
respectively. The latest version of the `Updated Nearby Galaxy Catalog'                           
(Karachentsev et al. 2013) contains 869 galaxies with distance estimates                          
$D<11$~Mpc.                                                                                       
                                                                                                  
In recent years, due to wide-field sky surveys in the optical range and in 
the \HI{} 21-cm line, the number of catalogued galaxies in the Local 
Volume (LV) with $D<11$~Mpc significantly increased, and for January 2018 it 
exeeded a thousand. The LV database\footnote{\url{http://www.sao.ru/lv/lvgdb}} 
contains different properties of nearby galaxies and atlas of their images 
(Kaisina et al. 2012). Galaxies have been included in the LV sample by 
one of two conditions: a) the radial velocity of a galaxy in reference 
to the Local Group (LG) centroid is $V_{LG}<600$~\kms{} or 
b) the estimate of its distance with any method is                         
$D<11$~Mpc. For $\sim400$ LV galaxies, distances were measured with an                            
accuracy of (5--10) per cent at the Hubble Space Telescope (HST) from the                         
tip of the red-giant branch (TRGB).                                                               
                                                                                                  
Currently, for most LV galaxies hydrogen masses are determined and                                
star-formation rates (SFR) are estimated by the flux in the far ultraviolet                       
(FUV), measured at the GALEX satellite, or in the \Halpha{} emission line.                        
Table 1 presents numbers of the LV galaxies observed in \HI{}, FUV and \Halpha{}                     
line as well as numbers of detected objects among them. Here we reduced the                       
total number of LV galaxies from 1153 to 1029, omitting 28 galaxies with                        
strong Galactic extinction $A_B > 3.0$ mag according to Schlafly \& Finkbeiner (2011)
and 98 galaxies which have been selected according to their low velocity $V_{LG}<600$~\kms{}                 
but whose distance estimates turn out then to be beyond 11 Mpc. The first row of                       
Table 1 indicates numbers of the LV galaxies belonging to different morphological                   
types on de Vaucouleurs scale: $T=10$ (Ir), $T=9$ (Im, BCD), $T=8-6$ (Sm, Sdm,                    
Sd, Sdc), $T=5-1$ (Sc,Sbc, Sb, Sab, Sa) and $T < 1$ (E, S0, dE, dSph).  

Appendix provides basic observation data on 1029 LV galaxies used in our analysis.
The full machine-readable version of the list is available at the Vizier 
database\footnote{\url{http://edsarc.u-strasbg.fr/viz-bin/qcalJ/other/}}. 
Individual references to original sources of the data can be found in the LV 
database.
                                                                        
 Abundant observables for this sample covering the stellar-mass range of                      
$\sim7$ dex and minamally burdened with selection effects allows one to study                     
different properties of galaxies depending on the environment in the wide                         
range of local densities ($\sim5$ dex within a sphere of the 1~Mpc radius).

\section{Indicators of environment density in the Local Volume} 

In the vicinity of every LV galaxy, there are numerous neighbours `n' with 
masses $M_n$ at spatial separations $D_n$. Due to enormous difference between 
galaxies by masses, the term `the nearest neighbour' loses its physical meaning; 
instead, the notion `the most significant neighbour' is advisable, the tidal 
force of which $F_n\sim M_n/D_n^3$ prevails the gravitational influence of other 
neighbours. We characterized each LV galaxy with the tidal index 
$$\Theta_1=\max[\log(M_n/D_n^3)]+C, n = 1, 2, ... N,$$ 
where the parameter $C=-10.96$ was chosen so that a galaxy with $\Theta_1=0$ 
was located on the `zero-velocity sphere' relative to its most significant 
neighbour, i.e., the Main Disturber (MD). In other words, a galaxy with 
$\Theta_1<0$ was considered a `field galaxy' causally unrelated to its MD, as 
the crossing time for this pair exceeds the age of the Universe.

However, the MD position could considerably vary with time owing to orbital 
motion, thus, we also used another, more robust, local-density indicator 
$$\Theta_5=\log(\sum^5_{n=1} M_n/D_n^3) +C,$$ 
which accounted for contribution of five most significant neighbours to 
density contrast. The $C$ parameter here is the same as in the case of $\Theta_1$. 

Finally, we have calculated the third indicator for each LV galaxy 
$$\Theta_j=\log[j_K(\mathrm{1~Mpc})/\langle j_K\rangle],$$
where  $j_K(\mathrm{1~Mpc})$ denotes  the average density of galaxy luminosity in the 
K-band  within a 1~Mpc radius and~$\langle j_K\rangle=4.28\times10^8L_{\odot}/$Mpc$^3$ 
(Jones et al. 2006) is the mean global density of the K-luminosity at the Hubble
parameter $H_0 = 73$~\kms{} Mpc$^{-1}$.
In so doing, the luminosity of the central galaxy itself was not taken into account.
If a galaxy had no neighbours within the 1~Mpc radius, its dimensionless parameter 
$\Theta_j$ was assumed equal to $-3.5$.

Hereinafter, we assume that the stellar mass of a galaxy is expressed through 
its luminosity in the K-band as 
$$M_*=1(M_{\odot}/L_{\odot})\times L_K$$
according to Bell et al. (2003). More recent estimates of the proportionality 
factor yield its value closer to $0.5(M_{\odot}/L_{\odot})$ 
(McGaugh \& Schombert 2014, Pomonareva et al. 2018) which should be kept in mind 
when normalizing the hydrogen mass $M_{\rm HI}$ and the star-formation rate SFR 
per unit $M_*$. The main source of data on K-band magnitudes is the 2MASS Sky Survey
(Jarrett et al. 2000, 2003) supplemented by photometric measurements from
(Fingerhut et al. 2010, Vaduvescu et al. 2006). At the lack of accurate 
photometry, K-magnitude was determined via apparent B- magnitude and morphological
type T by a relation (Jarrett et al. 2003): $\langle B - K\rangle = 4.10$ for $T<3$,
$\langle B - K\rangle = 2.35$ for $T>8$, and $\langle B - K\rangle = 4.60 -0.25 T$ for 
intermediate types $T =3-8$ by de Vaucouleurs scale.

The upper and lower panels in Fig.~1 show the relation between three tidal indices: 
$\Theta_1$ and $\Theta_5$, $\Theta_1$ and $\Theta_j$ for 1029 LV galaxies. 
As we can see, density variations determined by tidal indices $\Theta_1$ and 
$\Theta_5$ reach nine orders. Both the indices are closely correlated.
The density excess ($\Theta_5 - \Theta_1$) does not exceed 0.6 dex even in the 
most extreme cases. Therefore, further we do not use the parameter $\Theta_5$ in 
analysing various properties of LV galaxies.

As follows from the bottom panel of Fig.~1, the tidal index $\Theta_1$ and local 
density contrast $\Theta_j$ are also correlated, although, the statistical 
relation between them looks less clear and nonlinear.
The dispersion $\Theta_1$ grows with the increase of $\Theta_j$. 
The local-density range in a sphere of the 1~Mpc radius exceeds 5~dex.
To estimate specific features of this diagram, let us consider an extreme case 
of the ultra-compact companion SUCD1 (Hau et al. 2009) near the giant galaxy 
NGC~4594 = M104 = Sombrero. Physical connection of galaxies is confirmed by their 
radial velocities: 1109~\kms{} and 894~\kms{}, respectively.
The companion SUCD1 is located at a projected distance of $2.7^{\prime}$ from 
the center of Sombrero, which, with the TRGB distance of Sombrero of 9.55~Mpc 
(McQuinn et al. 2016), corresponds to 7.4~kpc.
The tidal index $\Theta_1$ for SUCD1 is calculated on the assumption of that 
the companion is located at the same distance as the giant galaxy.
If the distance to SUCD1 was measured with the TRGB method with the 5 per cent 
accuracy, then this would fix the mutual distance of two galaxies along the 
line of sight with an accuracy of $\pm480$~kpc. With such a separation shift 
by 480~kpc, the tidal index $\Theta_1 = 6.7$ could decrease to 1.3.
From this example, we deduce that a considerable part of the scatter of galaxies 
along the $\Theta_1$ axis is probably due to measurement errors of their distances.
Nevertheless, we preserve both parameters for further analysis: $\Theta_1$ and 
$\Theta_j$, giving a preference to the parameter $\Theta_1$ as more 
sensitive to SFR bursts and gas depletion in dwarf galaxies than the density 
contrast $\Theta_j$ in the 1~Mpc radius sphere.

\section{Gaseous fraction of mass vs. local density} 

The atomic hydrogen mass $M_{\rm HI}$ of a galaxy determined from its flux in 
the \HI{} line as 
$$(M_{\rm HI}/M_{\odot})=2.356\times10^5\times D^2\times F_{\rm HI}$$ 
is a parameter quite sensitive to the environment.
Here $D$ is in Mpc and the flux $F_{\rm HI}$ is in Jy\,\kms{}.
The effect of the \HI{} deficiency in spiral galaxies situated in groups and 
clusters was studied by Haynes \& Giovanelli (1984) in detail.
The sweeping-out of gas from galaxy disks during their motion through a dense 
virial core is assumed to be the main mechanism of formation of the \HI{} 
deficiency. Obviously, the sweeping-out of gas from dwarf galaxies happens 
easier due to a shallow potential well as compared to disks of massive galaxies.
Since the hydrogen mass correlates with the stellar mass of a galaxy and with 
its morphological type, then in order to reduce a scatter of data in diagrams, 
we consider the ratio $M_{\rm HI}/M_*$ for each morphological type separately.
An additional advantage arises under such an approach, as the ratio 
$M_{\rm HI}/M_*$ does not depend on errors of a galaxy distance.

Major sources of data on \HI{} fluxes of nearby galaxies are `blind' surveys 
of wide areas of sky with the radio telescopes: 
Parkes, HIPASS (Koribalski et al. 2004), Arecibo, ALFALFA (Haynes et al. 2011) 
and Westerbork, WSRT-CVn (Kovac et al. 2009).
Special observations of several hundred candidates to nearby dwarfs have been 
performed by Huchtmeier et al. (2000) at the 100-m dish in Effelsberg.
The data of second and third rows of Table 1 show that \HI{} fluxes are measured 
now for 794 objects among 1029 LV galaxies, but 210 of them have only upper 
limits of the flux. Unfortunately, a considerable area of the northern sky 
is still uncovered by a systematic \HI{}-survey.

Distributions of the LV galaxies with their hydrogen-to-stellar mass ratio 
depending on the tidal indices $\Theta_1$ and $\Theta_j$ are presented in a 
set of panels in Fig.~2, Fig.~3 and Fig.~4. A typical uncertainty of the 
\HI{}-mass fraction on the Figures is (0.1 - 0.2) dex. The galaxies with 
only upper limit of \HI{}-flux are shown by open circles.
For each morphological category, we calculated the 
linear-regression parameters $Y = a + bX$, the data on which are given in 
the upper part of Table~2. Here and so on, galaxies with upper limit
of their \HI{}, FUV or \Halpha{} fluxes were ignored. The Table contains 
regression parameters for both the tidal indices: $\Theta_1$ and $\Theta_j$.
The cases with a slope significant at the 3-sigma level are highlighted in 
Table 2 by boldface.
Considering the data of the Figures and Table~2, we notice the following 
features and trends.

\subsection{Dwarf irregular galaxies (T = 10)}

Despite the fact that irregular galaxies are assigned to one and the same type, 
T = 10, they show a great scatter of the hydrogen-to-stellar mass ratio.
Some dIr objects: ESO~215-009, And~IV, with $M_{\rm HI}/M_*>10$ have served as 
targets of a detailed study (Warren et al. 2004, Karachentsev et al. 2016). 
Both the gas-rich galaxies are isolated. However, the high ratio 
$M_{\rm HI}/M_*\simeq 15$ in another dwarf system, BK3N near M~81, seems to be 
caused by entangling \HI{} fluxes from these galaxies being in contact in the 
sky. It should be also noted that some low-surface-brightness dwarfs (e.g., 
HS~117) are referred to the dIr type (T = 10), but they look like transitional 
(Tr) between dIr and dSph by their smooth structure, reddish colour, and low 
hydrogen abundance.

On the whole, the dIr population demonstrates the decreasing ratio 
$M_{\rm HI}/M_*$ with the increase of the parameters $\Theta_1$ and 
$\Theta_j$. 
This effect is more prominent ($b=-0.060\pm0.021$) in the case of tidal 
index $\Theta_1$ that points on the primary role of the most significant 
neighbour (MD) in the process of gas evacuation from its dwarf companions.

As seen from Table 1, about 11 per cent of dIr galaxies observed in \HI{} have 
only an upper limit of their \HI{} flux (denoted with open circles in Fig.~2). 
Most of them are close companions to massive spirals having parameters 
$\Theta_1>1$ and $\Theta_j>1.5$.
Taking these galaxies into account would make the regression-line slope 
steeper.

The average hydrogen-to-stellar mass ratio in a population of irregular dwarfs 
is close to unity, i.e. they are in a half-way of their evolution.
The mean difference in $\log(M_{\rm HI}/M_*)$ from field galaxies to group 
members reaches 0.3~dex which is smaller than the dispersion of this ratio, 
$\sigma[\log(M_{\rm HI}/M_*)]\simeq 0.48$. 
Consequently, the environment density is not the basic factor determining the 
observed gas-to-stars fraction in irregular dwarfs.

\subsection{Magellanic and Blue Compact Dwarfs (T = 9)}

As seen from Table 2, the average stellar mass of these dwarfs is about 5 
times greater than that of dIrs, and the $M_{\rm HI}/M_*$ ratio is on 
average 2.5 times lower. The \HI{}-deficiency effect as a function of 
$\Theta_1$ or $\Theta_j$ is manifested as only a weak tendency with the 
regression line slope being within its statistical error.

The bright satellites of our Galaxy: the Large (LMC) and Small (SMC) Magellanic 
clouds, look like typical representatives of this population of dwarfs.
Wherein, there are examples of objects with the greater (AGC~112454) and 
very small (DDO~082) \HI{}-abundance.

\subsection{Bulgeless disks (T = 8 -- 6)} 

The last column in Table 2 shows that the typical stellar mass of late-type 
spirals in the Local Volume, $\langle\log M_*\rangle=9.14$, is much greater 
than that of irregular (7.46) and Magellanic (8.21) dwarfs. With the increase 
of the environment density, the \HI{}-deficiency effect is seen only as an 
insignificant tendency ($b=-0.02\pm0.02)$ of the expected sign.
All 130 LV galaxies of these morphological types are detected in \HI{}.
Relative abundance of hydrogen in them is almost the same ($-0.46$~dex) as 
that in BCD and Im dwarfs.

\subsection{Early-type spirals (T = 5 -- 1)} 

Galaxies of the Sc, Sb, and Sa types are inconsiderable in number in the Local 
Volume (N = 29). All of them are detected in \HI{}. Their average stellar mass 
is 10.63~dex, and the average hydrogen-to-stellar mass ratio is only $-1.44$~dex.
The $M_{\rm HI}/M_*$ ratio does not reveal any correlation with the environment 
parameters $\Theta_1$ and $\Theta_j$.

\subsection{E + S0 + dSph galaxies (T $<$ 1)} 

In this mixed sample of objects with old stellar population, dwarf spheroidal 
galaxies predominate in number. Current \HI{}-surveys are too shallow to 
reliably measure their \HI{} flux. As one can see from Table 1, in 93 per cent 
of cases (shown in bottom panel of Fig.4 by open circles), there was detected 
only the upper limit of \HI{}-flux. Most galaxies of this type are concentrated 
in high-density regions. In this regard, a specific selection effect arises: 
angular separations between spheroidal dwarfs and their massive neighbour are 
comparable to a beam size of radio telescopes which makes it difficult to measure 
the \HI{} flux from tight dwarf companions.

\section{Star-formation rate vs.\ local density}

Following Lee et al. (2009, 2011), we determined the integrated star-formation 
rate of a galaxy in units of $(M_{\odot}/$yr) as 
$$\log(\textrm{SFR})=2.78+2\log D-0.4\times m^c_{\rm FUV},$$
where the distance $D$ is expressed in Mpc and the apparent magnitude of a 
galaxy in the far ultraviolet, $m^c_{FUV}$, is measured at the GALEX satellite 
(Gil de Paz et al. 2007) and corrected for Galactic and internal extinction.
As it is shown in Table 1, the GALEX survey yields FUV fluxes for 647 LV 
galaxies and also flux upper-limits for other 230 galaxies.

In order to avoid errors of distance measurements and reduce the statistical 
scatter, we used the specific star-formation rate normalized to stellar mass 
of a galaxy $$\textrm{sSFR}=\textrm{SFR}/M_*.$$ 

Figs. 5--7 present the distribution of LV galaxies over the specific star-
formation rate and the environment-density indicators. A typical error of 
sSFR on them is (0.1 -- 0.2) dex. Galaxies were divided 
into morphological types in the same manner as in the previous section. The 
middle part of Table 2 shows the linear-regression parameters for each subsample.
Based on these data, we can notice the following tendencies.

\subsection{Dwarf irregular galaxies}

Among 339 LV dIr galaxies within the GALEX survey area, the FUV flux was 
detected in 301 objects, i.e., in almost 90 per cent of the sample. Their 
average star-formation rate is 
$$\langle\log(\textrm{sSFR})\rangle=-10.24 (yr^{-1}).$$ 
Consequently, with such a star-formation rate, being permanent throughout 
the age of the Universe, $T_0=10.14$ dex (yr), a typical irregular dwarf is 
able to reproduce 80 per cent of its stellar mass. (Note that assuming the ratio
$M_*=0.5(M_{\odot}/L_{\odot})\times L_K$ by McGaugh \& Schombert (2014) would 
yield the average star-formation rate for dIrs to be slightly higher in past 
than in the present epoch).

Passing from isolated irregular galaxies into group members, the average 
star-formation rate 3--4 times drops. This effect appears significant at the 
level of 4--5 sigma both by the parameter $\Theta_1$ with the regression slope 
of $b=-0.089\pm0.018$ and $\Theta_j$ with its slope of $b=-0.076\pm0.017$. 
Consequently, the presence of a massive neighbouring galaxy does not intensify, 
yet on average diminishes the star-formation rate of its dwarf companions.
This statement agrees with the result obtained earlier by Karachentsev et al.(2014)
but disagrees with conclusions derived by Ellison et al. 2008, Di Mateo et al.
2008, Lelli et al. 2014 and Knapen et al. 2015. In particular, Knapen et al. 2015 
investigated the influence of interactions on the star formation by studying a sample 
of 1478 nearby galaxies, all within a distance of 45 Mpc, and explored distribution 
of their sSFR with morphological type and with stellar mass. They found that sSFR is 
enhanced statistically in interacting galaxies. The increase is, however, moderate, 
reaching a maximum of a factor of 1.9 for the highest degree of interaction (i.e. 
mergers). A probable reason of the disagreement can be caused by different approaches
to the sample selection.  Knapen et al. 2015 selected preferably bright and large 
galaxies, while our LV sample consists of mainly low-mass galaxies esposed to quenching
action of a massive neighbor.

The distribution of irregular dwarfs by sSFR does not show a distinct lower limit.
Here one should also take into consideration that a part of the dIr population 
only have the upper estimate of their FUV flux. Furthermore, a number of dIrs (by our 
estimation, 40--50 per cent) have transitioned into the quenched dwarf spheroidal 
category for the time $T_0$. Along with this, a distinctive feature of this 
distribution is the presence of sharp upper limit (sSFR)$_{max}\simeq-9.4$~dex
that has been already noticed by (Karachentsev \& Kaisina 2013). The presence
of the upper limit can be also seen in observational data by Knapen et al.\ (2015), 
Pan et al.\ (2018), and in model calculations by Birrer et al.\ (2014). In our sample, 
there are only three irregular dwarfs near and slightly above this limit, the 
most outstanding of which is NGC~1592 with its ${\rm sSFR} = -9.15$~dex.
This isolated peculiar object looks like a chain of blue knots embedded in a 
faint envelope. Judging by a low color index $B - K = 1.55$ of NGC~1592, its 
K-luminosity from 2MASS survey (Jarrett et al. 2000, 2003), and accordingly the
stellar mass, seems to be about two times underestimated. Accounting for this 
correction drops NGC~1592 below the specified limit which presence we consider as 
important characteristic of the star-formation process in the current epoch.

\subsection{BCD and Im dwarfs}

Among 120 galaxies of this sample within the area of GALEX survey, only one 
object is undetected in the FUV band. The average star-formation rate of the 
BCD + Im dwarfs, $$\langle\log(\textrm{sSFR})\rangle=-10.21 (yr^{-1}),$$ is almost 
the same as that of irregular dwarfs. However, their sSFR dispersion (0.38) is  
appreciably smaller than that in the previous sample (0.49). These more massive 
dwarfs also show the effect of SFR decrease with the increase of the 
environment density, although, their regression slope is somewhat smaller: 
$b=-0.105\pm0.033$ for the parameter $\Theta_1$ and $b=-0.072\pm0.021$ for 
$\Theta_j$. 

Only one BCD dwarf, Mrk~36 = Haro~4, lies slightly above the limit 
$({\rm sSFR})_{\rm max} = -9.4$~dex. Probably, this moderately isolated object 
is in a short-term phase of burst activity, the cause of which, however, is 
not due to interaction with neighbours.

\subsection{Bulgeless disks} 

FUV fluxes are detected in all 118 late-type spiral galaxies observed with 
GALEX. Their average specific star-formation rate 
$\langle\log({\rm sSFR})\rangle=-10.14$ (yr$^{-1}$) just coincides with the 
Hubble parameter, $log(H_0)=-10.14$ (yr$^{-1}$). 
The star-formation rate in these galaxies is almost insensitive to the 
environment density. One can state that bulgeless galaxy disks have a 
uniform internal mechanism for converting gas into stars which works with 
remarkable consistency throughout the whole cosmological timescale 
$T_0=H_0^{-1}$.

\subsection{Early-type spirals} 

Representatives of this small subsample of LV galaxies significantly differ 
from each other in the disk-to-bulge mass ratio. For example, a disk is 
dominant in NGC~24, while in NGC~2787 and NGC~4594 (Sombrero) the main 
stellar mass is concentrated in their bulges. The GALEX survey detects the 
FUV flux in all 27 early-type spiral galaxies. However, the average specific 
star-formation rate for them, $\langle\log({\rm sSFR})\rangle=-10.95$, is about
six times lower than that for late-type spirals, since the old quenched population 
of bulges is involved in normalization per unit of the galaxy's stellar mass.
As was noticed by Abramson et al. 2014, normalizing to the mass of a disk 
itself, rather than to the total stellar galaxy mass, significantly reduces 
the dispersion of specific star-formation rates.

\subsection{Quenched E, S0, and dSph galaxies} 

As seen from Table 1, the depth of GALEX survey is too shallow to draw 
reliable conclusions about specific features of star formation in galaxies with 
an old stellar population. FUV fluxes are measured for only 82 of 273 LV 
galaxies of these types. As a rule, these are the nearest galaxies like M~32 
and NGC~205. Also, angular sizes for some dSphs in the Local Group (Fornax, 
Sculptor) are so large that estimation of their FUV flux can only be the 
upper limit due to abundant background objects falling into the galaxy borders.

The sSFR dispersion of the detected E, S0, and dSph galaxies is much greater 
than that of other types. The decreasing tendency of the star-formation rate of 
them with the increase of their environment density is weak. The average 
current star-formation rate for these galaxies is $-12.23$~dex (yr$^{-1}$) 
which is smaller than the $H_0$ cosmological scale by two orders of magnitude. 
According to Pipino et al. (2013), the global cosmic evolution of SFR as a 
function of a redshift Z splits into two different modes: a short epoch of 
intense star formation in E, S0, and dSph galaxies, as well in bulges of 
Sa--Sb galaxies before $Z=2$, and a stage of slow conversion of gas into 
stars in disks of late-type galaxies extended over the whole cosmic time $T_0$. 
This scenario quite agrees with observation data presented here for the LV 
galaxies.

\section{Star formation as seen in FUV and \Halpha{} surveys} 

Determination of SFR from the FUV flux corresponds to a characteristic 
timescale of $\sim100$ Myr. Another method uses the integrated flux from a 
galaxy in the \Halpha{} emission line. According to Kennicutt (1998), 
$$SFR = 0.945\times 10^9\times D^2\times F_c(H\alpha),$$ 
where the distance $D$ is expressed in Mpc and the flux in \Halpha{} in 
the units [erg$\times cm^{-2}\times s^{-1}$] is corrected for the Galactic 
(Schlafly \& Finkbeiner 2011) and internal extinction. This estimate 
characterizes the star-formation rate on a scale of $\sim10$~Myr. Thus, the 
ratio SFR(\Halpha{})/SFR(FUV) may be used to identify objects that are in 
a loud or quiet phases of their star-formation activity.

There are 573 galaxies in our sample with the estimated SFR(\Halpha{}).
Most of them (486) have the star-formation rate determined also via their FUV flux.
The majority of \Halpha{}-flux measurements were performed at the 6-m SAO RAS 
telescope with a typical error of $\sim0.07$~dex (see Kaisin \& Karachentsev 
2014 and references therein).

The distribution of LV galaxies by the SFR(\Halpha{})/SFR(FUV) ratio and the 
environment-density indicator $\Theta_1$ is given in the panels of Figs.~8. 
Objects with the upper limits for the F(\Halpha{}) and F(FUV) fluxes are 
represented by point-down or point-up triangles, respectively. Galaxies are 
divided by morphological type in the same manner as before. Linear-regression 
parameters for them are shown at the bottom of Table~2. Considering these data, 
we can draw the following tendencies: 

a) The average value of log [SFR(\Halpha{})/SFR(FUV)] monotonously increases 
along the Hubble sequence from T = 10 to T = 1. The amplitude of this variation 
reaches a value of 0.52~dex.

b) The smallest scatter of SFR(\Halpha{})/SFR(FUV) ratio is characteristic 
of late-type bulgeless spiral galaxies (T = 8--6).

c) The trends of the SFR(\Halpha{})/SFR(FUV) versus environment density
look indistinct for galaxies of all types. Only in the case of {E, S0, dSph}- 
galaxies, the ratio decreases from isolated galaxies to group members with 
the slope $b=-0.268\pm0.085$ significant at the 3-sigma level.

Note that among extreme deviations of SFR(\Halpha{})/SFR(FUV) $>10$ there are 
both the cases of a galaxy being in the short-term phase of burst activity 
(M~82) and also the cases of erroneous overestimation of the \Halpha{} flux: 
NGC~6503-d1 (Koda et al. 2015), JKB83 (James et al. 2017). On the other side, 
galaxies with SFR(\Halpha{})/SFR(FUV) $<1/10$, for example, KDG~52 and DDO~120, 
are apparently at a calm stage between bursts.

Sperello di Serego Alighieri has drawn our attention to the fact that neutral 
hydrogen in galaxies can be ionized not only by young stars but also by hot 
evolved stars. Therefore, in galaxies, where the old stellar population is 
dominant, the \Halpha{} flux is not closely related to the star-formation 
process (Trinchieri \& de Serego Alighieri 1991, Binette et al. 1994).
Accounting for this factor is able to explain the high SFR(\Halpha{})/SFR(FUV) 
ratio which occurs for the E, S0 galaxies like NGC~4600 and for bulges of the 
S0a galaxies like NGC~3593.

Variations of the SFR(\Halpha{})/SFR(FUV) ratio depending on the SFR(FUV) value, 
or galaxy mass, or morphological type were discussed by many authors 
(Lee et al. 2009, Meurer et al. 2009, Hunter et al. 2010, Fumagalli et al. 2011, 
Weisz et al. 2012, Karachentsev \& Kaisina 2013, Lee et al. 2016, Watkins 
et al. 2017, Shimakawa et al. 2017). Pflamm-Altenburg et al. (2007, 2009) 
noticed that the observed systematic decrease of SFR(\Halpha{}) relative to 
SFR(FUV) from massive galaxies towards dwarfs could be caused by underestimating 
SFR via the \Halpha{} flux due to shortage of the brightest young stars in 
dwarf galaxies. Fumagalli et al. (2011) and Weisz et al. (2012) suppose that 
the reason for this difference is mainly the bursty character of star formation 
in low-mass dwarfs.

The panels of Fig.~9 show the SFR(\Halpha{})/SFR(FUV) ratio as the stellar-mass 
function for various morphological types. Linear-regression parameters for them 
are shown at the bottom of Table~2. As seen from these data, the average 
value of $\log M_*$ grows along the Hubble sequence from T = 10 to T = 5--1 by 
three orders of magnitude. With the increase of $M_*$, the ratio 
SFR(\Halpha{})/SFR(FUV) grows by 0.5~dex. The smallest dispersion of this ratio, 
0.298, is characteristic of the disks of spiral bulgeless galaxies which agrees 
with the concept of the decreasing role of star-formation bursts in the 
transition from dwarfs to massive galaxies.

Kaisin \& Karachentsev (2014) imaged in the \Halpha{} line a hundred LV 
galaxies without known FUV fluxes. Based on the obtained data, SFR(\Halpha{}) 
estimates for them can be converted to the SFR (FUV) system via the 
empirical relation 
$$ \langle\log[SFR(H\alpha)/SFR(FUV)]\rangle = 0.16\times\log M_*-1.58.$$ 
On average, for Im and BCD galaxies, this relation yields a small correction 
equal to $-0.20$~dex, while for dIr galaxies, the correction is $-0.40$~dex.
As seen from Fig.~9, data statistics for the {E, S0, dSph}- galaxies is so 
poor that it disallows any quantitative estimation.

When determining the SFR(\Halpha{}) and SFR(FUV) fluxes, we took into 
account the internal extinction in a galaxy: 
$$A(H\alpha)=0.54\times A_B \,\,\,\, \& \,\,\,\, A(FUV)=1.93\times A_B,$$ 
where $A_B$ is the extinction in the B band (in mag).
According to Verheijen (2001), $A_B$ can be expressed through the apparent 
axial ratio of a galaxy, a/b, and the rotation amplitude $V_m$ (in \kms{}) as 
$$A_B=[1.54+2.54(\log V_m-2.2)]\log(a/b)$$ 
if $V_m>40$~\kms{} and $A_B=0$ for dwarfs with smaller rotation amplitude.
Accounting for this relation, we built the diagrams of 
log[SFR(\Halpha{})/SFR(FUV)] vs.\ $\log(a/b)$ for irregular galaxies (T = 10, 9), 
late-type (T = 8--6), and early-type (T = 5--1) spiral galaxies.
The panels of Fig.~10 show the results, while the linear-regression parameters 
are given in the three last rows of Table~2. Dwarf irregular galaxies as well as 
early-type spirals do not exhibit any significant trend. But for the late-type 
spirals, such a trend is seen and described by the regression 
$$\log[SFR(H\alpha)/SFR(FUV)]=-(0.24\pm0.11)\log(a/b)-0.06.$$ 
Taking into consideration the relations between $A(H\alpha), A(FUV)$, 
and $A_B$, we can conclude that the recipe from Verheijen (2001) 
overestimates the internal extinction in Scd, Sd, and Sm galaxies by 
the value $\Delta A_B=(0.43^m\pm0.20^m)\log(a/b)$. Such a correction may 
be essential for thin edge-on galaxies.

\section{Morphological type and surface brightness vs.\ environment} 

The parameters considered above, $M_{\rm HI}/M_*$ and $SFR/M_*$, are rather 
sensitive to the environment density, since the light gaseous component of a 
galaxy is easily exposed to external influence. This is especially true for 
low-mass dwarf systems that dominate in the LV sample. Suppressing or 
intensifying star formation in a galaxy, the external influence affects also 
the morphological type and surface brightness of a galaxy.

Two panels of Fig.~11 show the dependence between the relative number of 
quenched {E, S0, dSph}- galaxies and the environment density parameters 
$\Theta_1$ and $\Theta_j$. The vertical bars correspond to the standard error 
of the mean. The bottom panel shows that among field galaxies with $\Theta_j<1$, 
the fraction of passive galaxies is only 0--10 per cent, but in dense regions 
with $\Theta_j>1$ their abundance sharply increases to 40--50 per cent. 
The top panel demonstrates a somewhat different pattern. The local density 
contrast $\Theta_1$ from the most significant neighbour shapes a monotonous 
growth of f(E) from $\sim0$ to $\sim50$ per cent at $\Theta_1<2$. However, 
there is not any further growth seen at higher $\Theta_1$ (possibly due to 
galaxy distance errors comparable to the virial radius of a group). Similar 
behaviour of a quenched fraction of satellites around the nearby luminous galaxies 
has been also noticed by Karachentsev \& Kudrya (2015) and Fillingham et al. (2018).

As follows from the data in Fig.~12, the average surface brightness of galaxies 
within their Holmberg isophote in the B band varies by nearly one magnitude 
depending on the environment density. This effect is appreciably larger than 
statistical errors indicated by vertical bars. The average surface brightness is 
almost constant for field galaxies with $\Theta_j<1$, while galaxies in groups 
with $\Theta_j>1$ appear fainter, probably due to the presence of a considerable 
number of quenched spheroidal dwarfs. In the top diagram, the average surface 
brightness also drops with the growth of parameter $\Theta_1$, but it shows 
the reverse tendency with $\Theta_1>4$. Possibly this reversal is caused by 
tidal stripping of stellar periphery of companions located very closely to a 
massive neighbour. The SUCD1 ultra-compact companion near the massive Sombrero 
galaxy serves as an example.

Note, that theoretical predictions about the behaviour of surface brightness 
of galaxies in various environments are still rather sparse and vague.
According to Hearin et al. (2017), `small galaxies cluster much more strongly 
than large galaxies of the same stellar mass', but they have not found any 
quantitative estimate of this effect in terms of surface brightness.

\section{Concluding remarks} 

We have considered influence of an environment on global properties of galaxies: 
the neutral hydrogen abundance, star-formation rate, morphological type, and 
average surface brightness. The sample used contains 1029 galaxies within the 
sphere of 11~Mpc radius around the Milky Way. The bulk of this sample, 
$\sim85$per cent, is composed of dwarf galaxies most exposed to external action 
due to their shallow potential well.

To characterize an environment, we used two parameters: the density contrast 
$\Theta_1$ provided by the most significant neighbour, and the local density 
excess $\Theta_j$ within 1-Mpc distance around a galaxy taken relative to the 
average cosmic density. Variations in the values of the parameters $\Theta_1$ 
and $\Theta_j$ exceed five orders of magnitude. The hydrogen-to-stellar mass 
ratio of a galaxy reveals a weak effect of the \HI{} deficiency in high-density 
regions being most noticeable in the case of low-mass irregular dwarfs.

The star-formation rate of the LV galaxies normalized to a unit of stellar mass 
turns out to be more sensitive to the environment density than $M_{\rm HI}/M_*$ 
ratio. Again, the fall of the specific star-formation rate with the growth of 
the environment density is most significant for the Ir, Im, and BCD dwarfs.

With a permanent present star-formation rate, dwarf galaxies and spiral galaxies 
without massive bulges are able to reproduce in average their observed stellar 
mass over the cosmological time $T_0=10.14$~dex (yr). Spiral galaxies of early 
types (Sa--Sb) and E, S0, dSph galaxies have the present sSFR able to reproduce 
only $\sim(1\textrm{--}10)$ per cent of their stellar mass for the time $T_0$. 
In other words, formation of stellar population in E, S0, and Sa galaxies has 
a significantly different (short and stormy) history than the sluggish process 
of converting gas into stars in irregular galaxies and spiral disks.

More than 99 per cent of the LV galaxies have a specific star-formation rate 
below $-9.4$~dex (yr$^{-1}$) i.e. below $5.5\times H_0$. This critical 
threshold seems to be an important characteristic governing the evolution of 
the interstellar medium in the current epoch.

Comparison between the sSFRs determined from the \Halpha{} and FUV fluxes 
shows a good mutual calibration for disks of spiral (T = 1--8) galaxies, 
being only slightly dependent on the environment. However, for dwarfs of the
dIr and Im+BCD types, the average displacement of log[SFR(\Halpha{})/SFR(FUV)] 
is $-0.40$~dex and $-0.20$~dex, respectively. Also, the dispersion of this 
ratio significantly increases from spiral disks towards irregular dwarfs which 
is apparently indicative of star-formation bursts in dwarfs on a scale of 
$\sim10^8$~yr (Skillman 2005, Stinson et al. 2007, Di Matteo et al. 2008).

The average SFR(\Halpha{})/SFR(FUV) ratio decreases with the increase of  
apparent axial ratio of a galaxy which is especially noticeable for thin 
Scd--Sd-Sdm disks. This effect is evidently due to usual overestimation of 
the internal extinction in galaxy disks via the recipe by Verheijen (2001). 
For thin disks with the axial ratio $a/b = 10$, the extinction excess in the 
B-band amounts to $-0.43^m\pm0.20^m$. 

The abundance of quenched galaxies of the E, S0, dSph types grows from 
0--10 per cent among field galaxies to 50 per cent in high-density regions, 
which is a demonstration of the well-known effect of the morphological 
segregation of galaxies. The average surface brightness of the LV galaxies 
also significantly varies with the environment density: dimmer galaxies 
without signs of star formation are more frequent in dense regions than in 
the general field. However, in the tight proximity of massive galaxies, one 
can see some indications of a competitive mechanism -- stripping stellar 
periphery of dwarf companions by the tidal force from a massive neighbour.

Finally, we need to notice that about 20--40 per cent of the LV galaxies still 
do not have flux estimates both in the $\HI{}$ and in \Halpha{} emission lines, 
and in the FUV band. Increasing the completeness of these surveys will allow 
one to take into consideration various effects of observational selection 
responsible for statistical biases more properly.

We thank the anonymous referee for numerous helpful comments that 
essentially improved the paper. This work is supported by the RSF grant no.\ 14--12--00965.
The update of the database on LV galaxies is supported by the RFBR grant no.\ 18--02--00005.

\label{lastpage}

\clearpage
\begin{table} 
\caption{Numbers of UNGC galaxies observed and detected in \HI{}, FUV, \Halpha{}.}
\begin{tabular}{lrrrrrr}
\hline\hline
Sample numbers &All types  &T=10 &T=9  &T=8-6 &T=5-1 &T<1\\
\hline
UNGC (D<11.0,A<3.0)                               &1029     &      395  &     148  &     138    &     29   &     319 \\       
Observed in \HI{}                                 & 794     &      329  &     127  &     130    &     29   &     179 \\
Detected in \HI{}                                 & 584     &      293  &     119  &     130    &     29   &      13 \\
Observed in FUV                                   & 877     &      339  &     120  &     118    &     27   &     273 \\
Detected in FUV                                   & 647     &      301  &     119  &     118    &     27   &      82 \\
Observed in \Halpha{}                             & 573     &      228  &     89   &     117    &     29   &     110 \\    
Detected in \Halpha{}                             & 470     &      195  &     84   &     113    &     29   &      49 \\ 
Detected both in \HI{} and FUV or \Halpha{}       & 557     &      276  &     111  &     127    &     29   &      12 \\  
\hline\hline								       
\end{tabular} 
\end{table}

\begin{table} 
\caption{Linear regression parameters $Y = a + b X$ for the LV galaxies.} 
\begin{tabular}{lccrrrrr} 
\hline\hline
Y &X &T &N &b$\pm\sigma_b$ &$\sigma_Y$ &$\langle Y \rangle$ &$\langle X \rangle$\\
\hline
log($M_{\rm HI}/M_*$)   &$\Theta_1$ &10   &293 &$\mathbf{-0.060\pm0.020}$ &0.475 &$-0.060 $&$-0.043$\\
log($M_{\rm HI}/M_*$)   &$\Theta_j$ &10   &293 &$-0.025\pm0.016$          &0.479 &$       $&$ 0.033$\\
log($M_{\rm HI}/M_*$)   &$\Theta_1$ &9    &119 &$-0.016\pm0.040$          &0.494 &$-0.475 $&$-0.211$\\
log($M_{\rm HI}/M_*$)   &$\Theta_j$ &9    &119 &$-0.015\pm0.026$          &0.493 &$       $&$ 0.004$\\
log($M_{\rm HI}/M_*$)   &$\Theta_1$ &8--6 &130 &$-0.020\pm0.027$          &0.439 &$-0.461 $&$ 0.280$\\
log($M_{\rm HI}/M_*$)   &$\Theta_j$ &8--6 &130 &$-0.022\pm0.025$          &0.438 &$       $&$ 0.141$\\
log($M_{\rm HI}/M_*$)   &$\Theta_1$ &5--1 &29  &$ 0.024\pm0.092$          &0.603 &$-1.440 $&$ 0.546$\\
log($M_{\rm HI}/M_*$)   &$\Theta_j$ &5--1 &29  &$-0.001\pm0.117$          &0.604 &$       $&$ 0.442$\\
log($M_{\rm HI}/M_*$)   &$\Theta_1$ &$<$1 &13  &$-0.229\pm0.152$          &0.907 &$-1.865 $&$ 0.926$\\
log($M_{\rm HI}/M_*$)   &$\Theta_j$ &$<$1 &13  &$-0.256\pm0.152$          &0.888 &$       $&$ 0.554$\\
log(sSFR$_{FUV}$)       &$\Theta_1$ &10   &301 &$\mathbf{-0.089\pm0.018}$ &0.487 &$-10.239$&$ 0.356$\\
log(sSFR$_{FUV}$)       &$\Theta_j$ &10   &301 &$\mathbf{-0.076\pm0.017}$ &0.490 &$       $&$ 0.348$\\
log(sSFR$_{FUV}$)       &$\Theta_1$ &9    &119 &$\mathbf{-0.105\pm0.033}$ &0.380 &$-10.212$&$-0.163$\\
log(sSFR$_{FUV}$)       &$\Theta_j$ &9    &119 &$\mathbf{-0.072\pm0.021}$ &0.378 &$       $&$ 0.198$\\
log(sSFR$_{FUV}$)       &$\Theta_1$ &8--6 &118 &$-0.002\pm0.024$          &0.381 &$-10.135$&$ 0.353$\\
log(sSFR$_{FUV}$)       &$\Theta_j$ &8--6 &118 &$-0.051\pm0.025$          &0.374 &$       $&$ 0.241$\\
log(sSFR$_{FUV}$)       &$\Theta_1$ &5--1 &27  &$ 0.084\pm0.092$          &0.600 &$-10.950$&$ 0.515$\\
log(sSFR$_{FUV}$)       &$\Theta_j$ &5--1 &27  &$ 0.094\pm0.121$          &0.603 &$       $&$ 0.397$\\
log(sSFR$_{FUV}$)       &$\Theta_1$ &$<$1 &82  &$-0.106\pm0.054$          &0.831 &$-12.227$&$ 1.856$\\
log(sSFR$_{FUV}$)       &$\Theta_j$ &$<$1 &82  &$-0.152\pm0.082$          &0.833 &$       $&$ 1.293$\\
log(SFR[\Halpha{}/FUV]) &$\Theta_1$ &10   &161 &$ 0.054\pm0.030$          &0.535 &$-0.398 $&$ 0.097$\\
log(SFR[\Halpha{}/FUV]) &$\Theta_1$ &9    &68  &$-0.081\pm0.052$          &0.461 &$-0.204 $&$-0.240$\\
log(SFR[\Halpha{}/FUV]) &$\Theta_1$ &8--6 &98  &$ 0.018\pm0.021$          &0.300 &$-0.149 $&$ 0.425$\\
log(SFR[\Halpha{}/FUV]) &$\Theta_1$ &5--1 &27  &$-0.085\pm0.047$          &0.308 &$ 0.118 $&$ 0.515$\\
log(SFR[\Halpha{}/FUV]) &$\Theta_1$ &$<$1 &23  &$\mathbf{-0.268\pm0.085}$ &0.696 &$ 0.011 $&$ 1.700$\\
log(SFR[\Halpha{}/FUV]) &$\Theta_j$ &10   &161 &$-0.005\pm0.026$          &0.540 &$-0.398 $&$ 0.220$\\
log(SFR[\Halpha{}/FUV]) &$\Theta_j$ &9    &68  &$-0.053\pm0.033$          &0.461 &$-0.204 $&$-0.089$\\
log(SFR[\Halpha{}/FUV]) &$\Theta_j$ &8--6 &98  &$ 0.030\pm0.023$          &0.298 &$-0.149 $&$ 0.318$\\
log(SFR[\Halpha{}/FUV]) &$\Theta_j$ &5--1 &27  &$-0.001\pm0.066$          &0.327 &$ 0.118 $&$ 0.397$\\
log(SFR[\Halpha{}/FUV]) &$\Theta_j$ &$<$1 &23  &$-0.113\pm0.139$          &0.830 &$ 0.011 $&$ 1.090$\\
log(SFR[\Halpha{}/FUV]) &log($M_*$) &10   &161 &$ 0.129\pm0.068$          &0.534 &$-0.398 $&$ 7.463$\\
log(SFR[\Halpha{}/FUV]) &log($M_*$) &9    &68  &$ 0.142\pm0.102$          &0.463 &$-0.204 $&$ 8.209$\\
log(SFR[\Halpha{}/FUV]) &log($M_*$) &8--6 &98  &$ 0.117\pm0.041$          &0.289 &$-0.149 $&$ 9.143$\\
log(SFR[\Halpha{}/FUV]) &log($M_*$) &5--1 &27  &$-0.018\pm0.132$          &0.327 &$ 0.118 $&$10.627$\\
log(SFR[\Halpha{}/FUV]) &log($M_*$) &$<$1 &23  &$ 0.074\pm0.102$          &0.832 &$ 0.011 $&$ 8.173$\\
log(SFR[\Halpha{}/FUV]) &log(a/b)   &10--9&229 &$-0.279\pm0.263$          &0.525 &$-0.340 $&$ 0.220$\\
log[SFR(\Halpha{}/FUV]) &log(a/b)   &8--6 &98  &$-0.237\pm0.107$          &0.294 &$-0.149 $&$ 0.382$\\
log[SFR(\Halpha{}/FUV]) &log(a/b)   &5--1 &27  &$-0.210\pm0.321$          &0.324 &$ 0.118 $&$ 0.282$\\
\hline\hline
\end{tabular} 
\end{table} 

\clearpage

\begin{figure*} 
\includegraphics[scale=1.2]{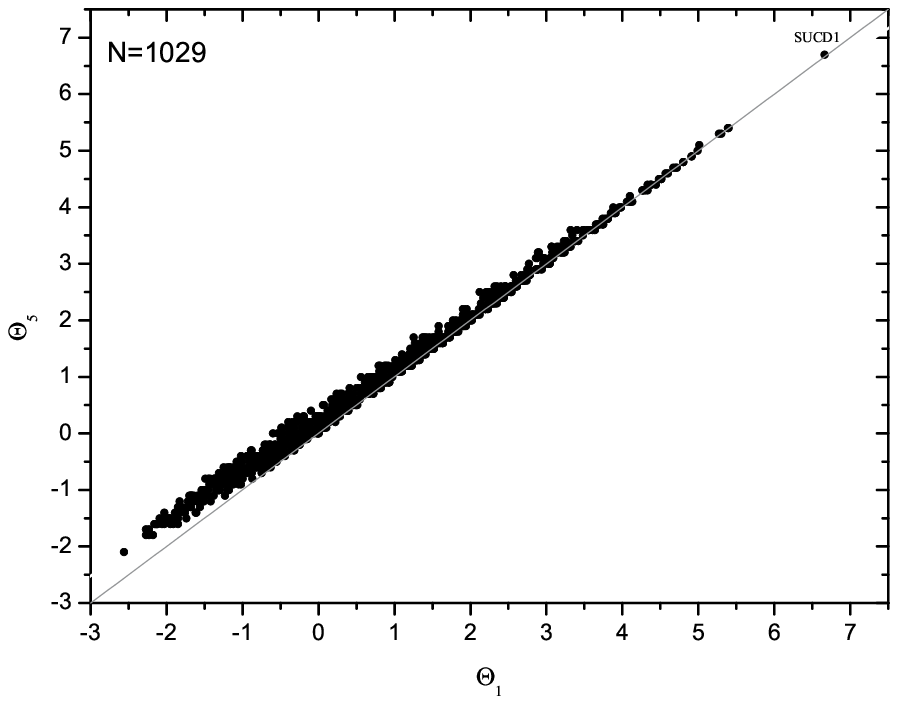}\\ \includegraphics[scale=1.2]{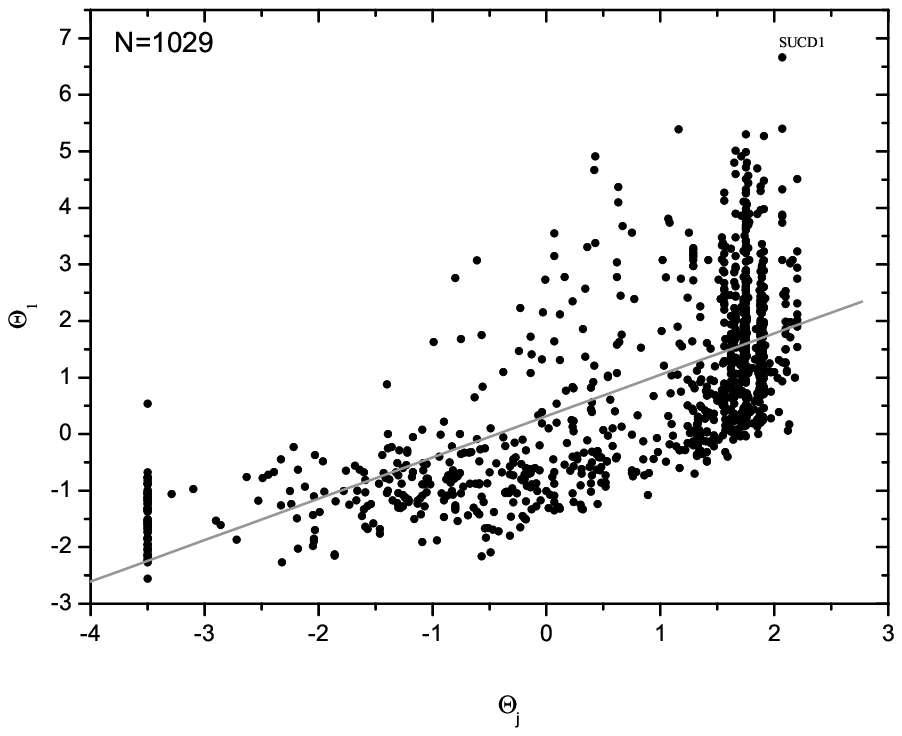} 
\caption{Comparison between three indices of the local density contrast.
The upper panel -- the tidal index determined by five significant neighbours versus the tidal index determined by the most significant neighbour.
Lower panel -- the tidal index determined by the most significant neighbour versus the local density contrast within 1 Mpc.} 
\end{figure*} 

\begin{figure*} 
\includegraphics[scale=1.2]{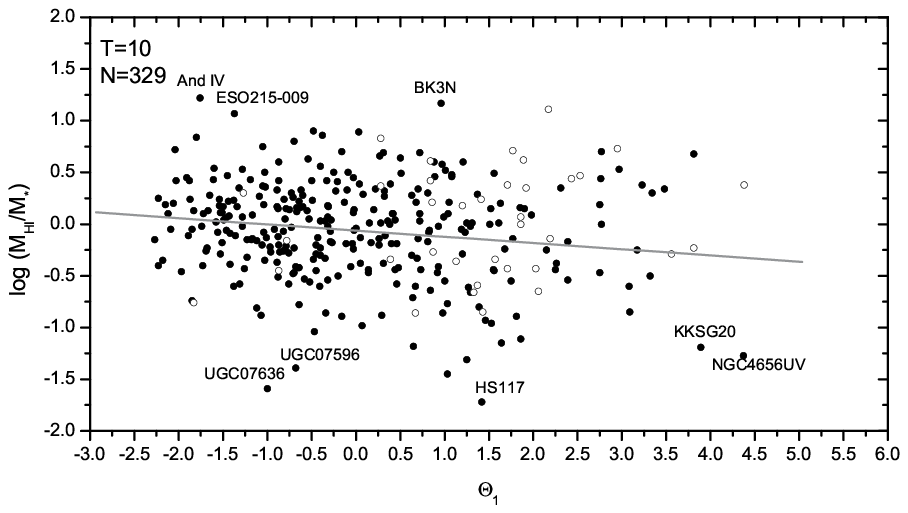}\\ \includegraphics[scale=1.2]{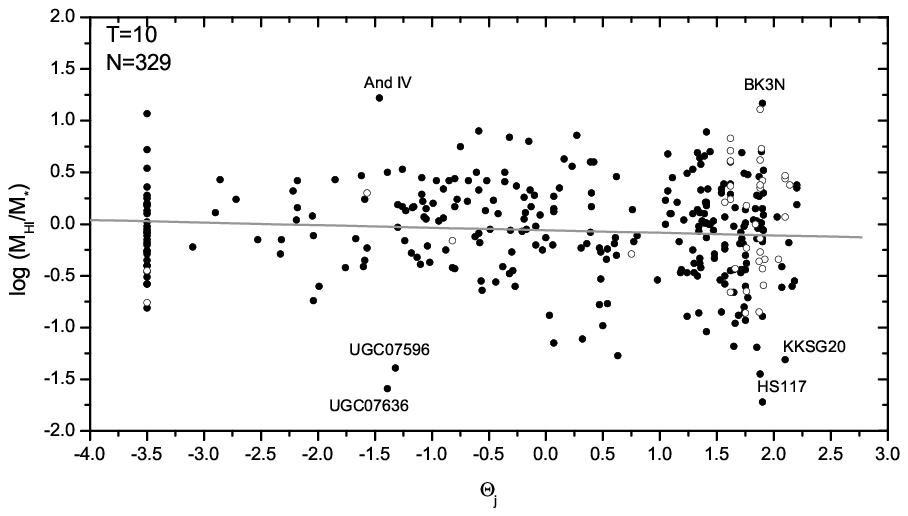} 
\caption{Hydrogen-to-stellar mass ratio as a function of the tidal index $\Theta_1$ (the upper panel) and the local density contrast $\Theta_j$ (the lower panel) for dIr (T = 10) galaxies. The galaxies with the upper limit of \HI{} flux are shown by open circles. A typical uncertainty of the \HI{} mass fraction 
on this and two next Figures is (0.1 - 0.2) dex.} 
\end{figure*} 

\begin{figure*} 
\includegraphics[scale=1.2]{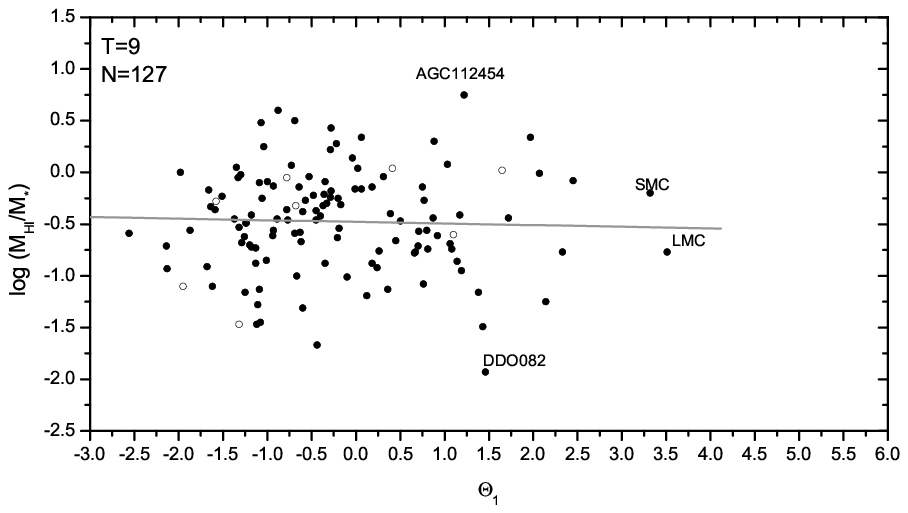}\\ \includegraphics[scale=1.2]{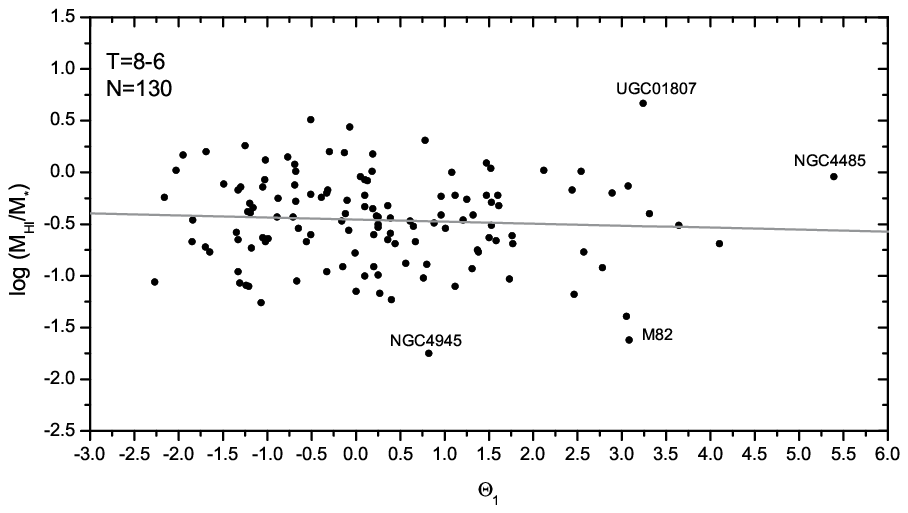} 
\caption{Hydrogen-to-stellar mass ratio as a function of the tidal index $\Theta_1$  for Magellanic and Blue Compact Dwarfs (T = 9) (the upper panel) and  for late-type spirals (T = 8 -- 6) (the lower panel).} 
\end{figure*} 

\begin{figure*} 
\includegraphics[scale=1.2]{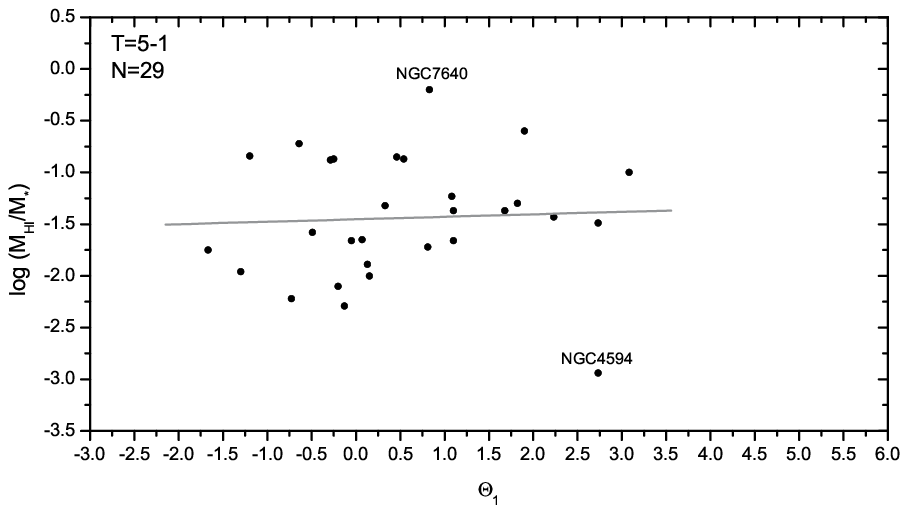}\\ \includegraphics[scale=1.2]{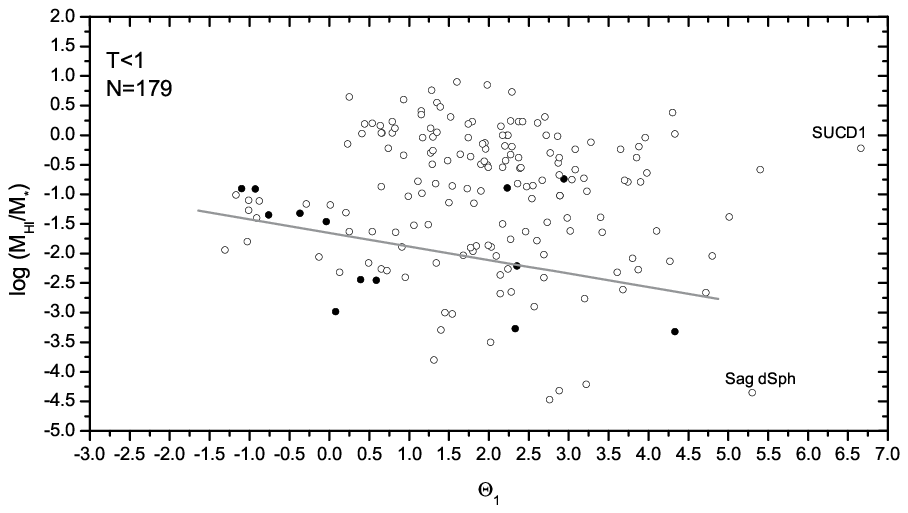} 
\caption{$M_{\rm HI}/M_*$ ratio as a function of $\Theta_1$ for early-type spiral galaxies (T = 5 -- 1) (the upper panel) and for E, S0, and dSph galaxies (the lower panel). The galaxies with the upper limit of \HI{} flux are shown by open circles and not accounted for the regression line.} 
\end{figure*} 

\begin{figure*} 
\includegraphics[scale=1.2]{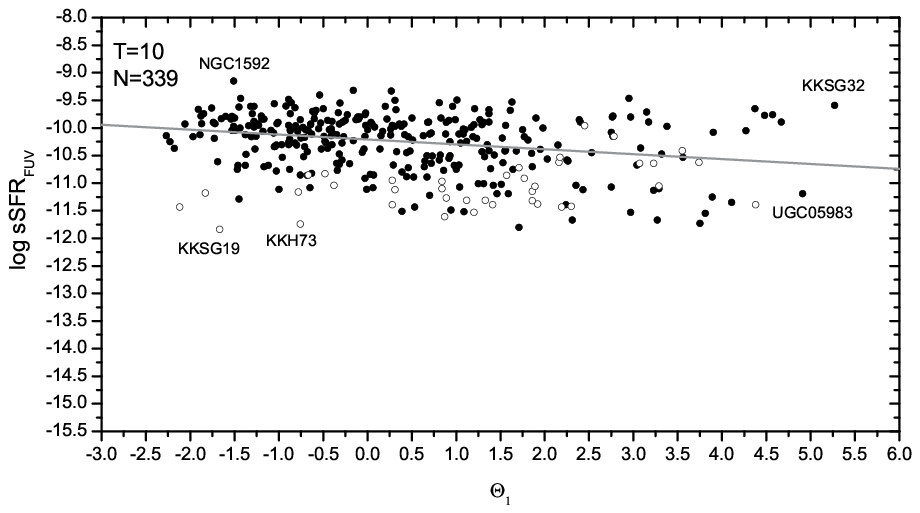}\\ \includegraphics[scale=1.2]{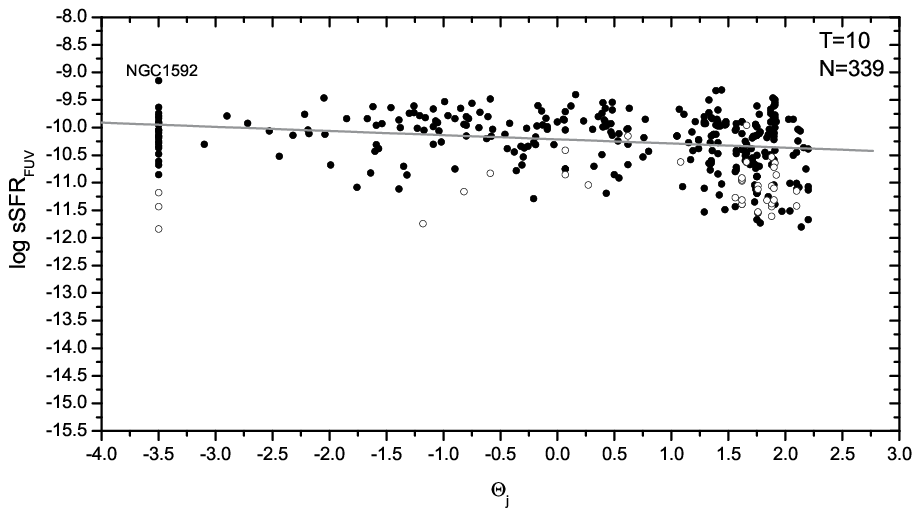} 
\caption{Specific star-formation rate versus $\Theta_1$ (the upper panel) and $\Theta_j$ (the lower panel) for dIr galaxies. The galaxies with the upper FUV flux limit are shown by open circles. A typical uncertainty of sSFR on this and two next Figures is (0.1 - 0.2) dex.} 
\end{figure*} 

\begin{figure*} 
\includegraphics[scale=1.2]{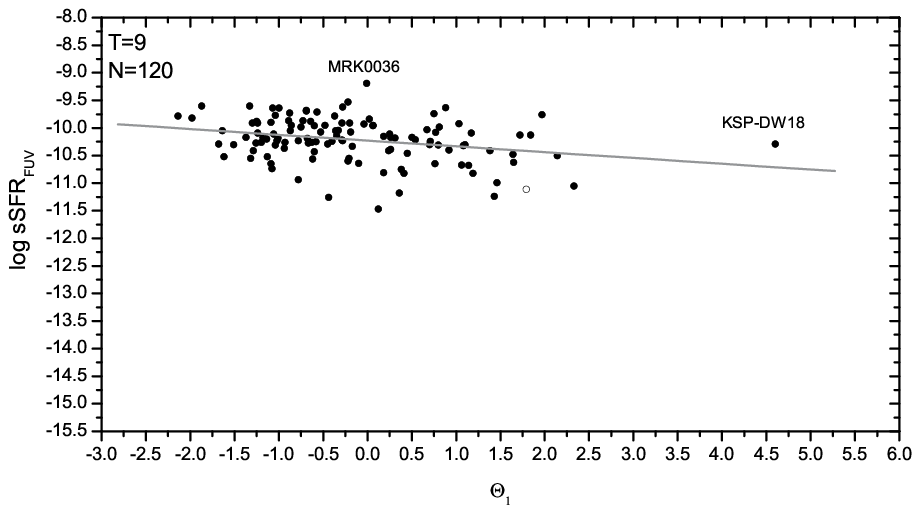}\\ \includegraphics[scale=1.2]{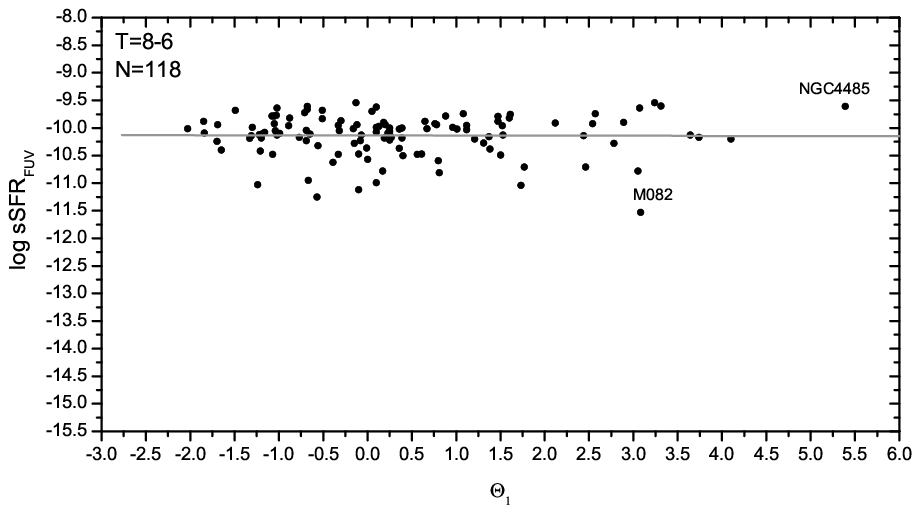} 
\caption{sSFR versus $\Theta_1$ for Magellanic and Blue Compact Dwarfs (the upper panel) and for late-type spirals (the lower panel).} 
\end{figure*} 

\begin{figure*} 
\includegraphics[scale=1.2]{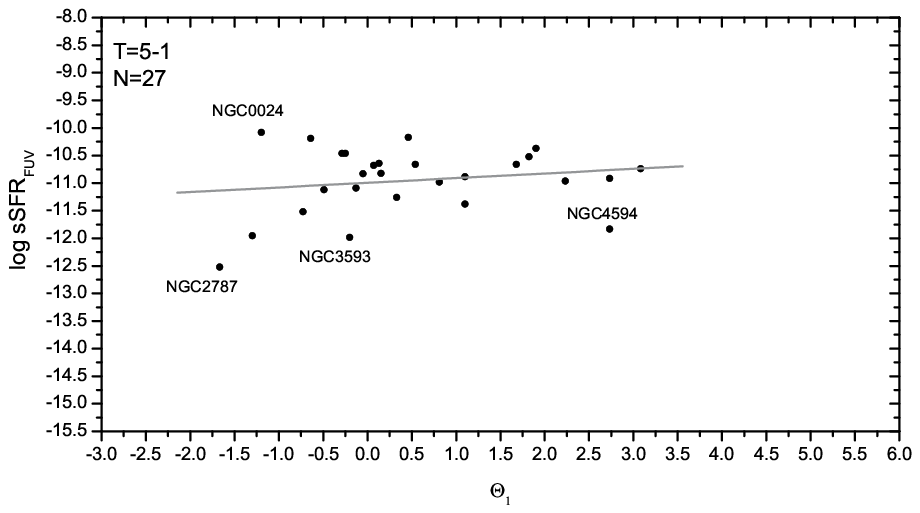}\\ \includegraphics[scale=1.2]{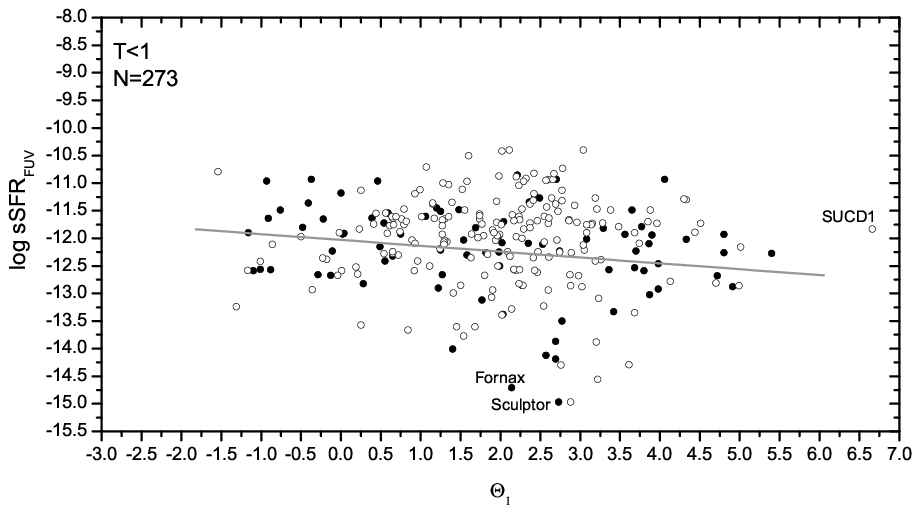} 
\caption{sSFR versus $\Theta_1$ for early-type spiral galaxies (the upper panel) and for E, S0, and dSph galaxies (the lower panel).} 
\end{figure*} 


\begin{figure*} 
\includegraphics[scale=0.08]{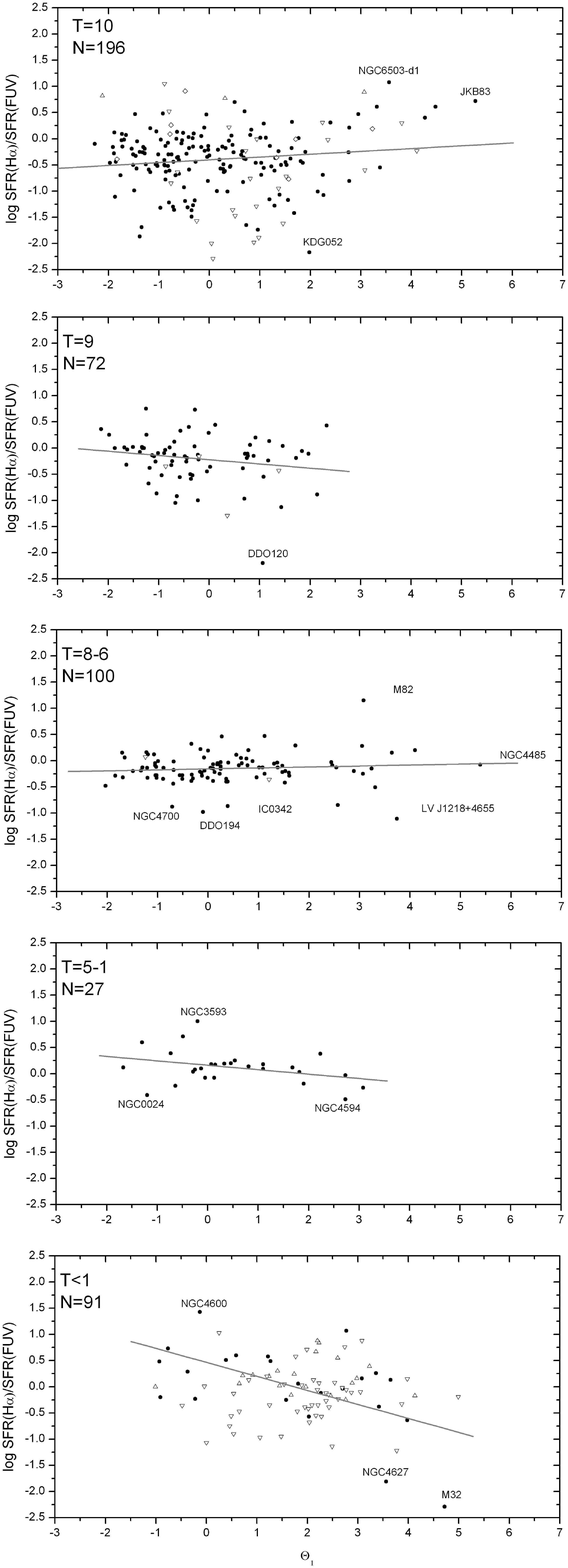} 
\caption{SFR (H$\alpha$)/SFR(FUV) versus $\Theta_1$ for the LV galaxies of different morphological types.
The galaxies with the upper limit to their \Halpha{} or FUV flux are shown by point-down or point-up open triangles, respectively
and not accounted for the regression line.} 
\end{figure*} 

\begin{figure*} 
\includegraphics[scale=0.08]{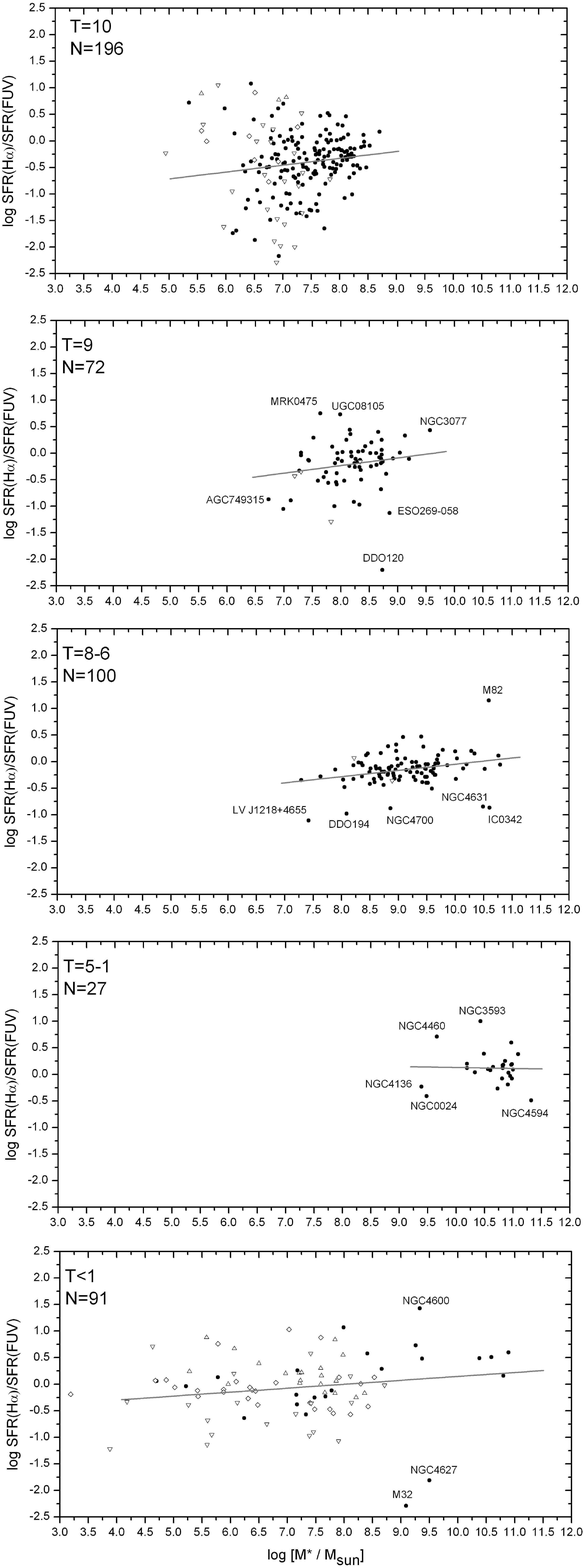} 
\caption{SFR (\Halpha{})/SFR(FUV) versus the stellar mass for the LV galaxies of different morphological types.
The galaxies with the upper limit to their \Halpha{} or FUV flux are shown by point-down or point-up open triangles, respectively.} 
\end{figure*} 

\begin{figure*} 
\includegraphics[scale=0.1]{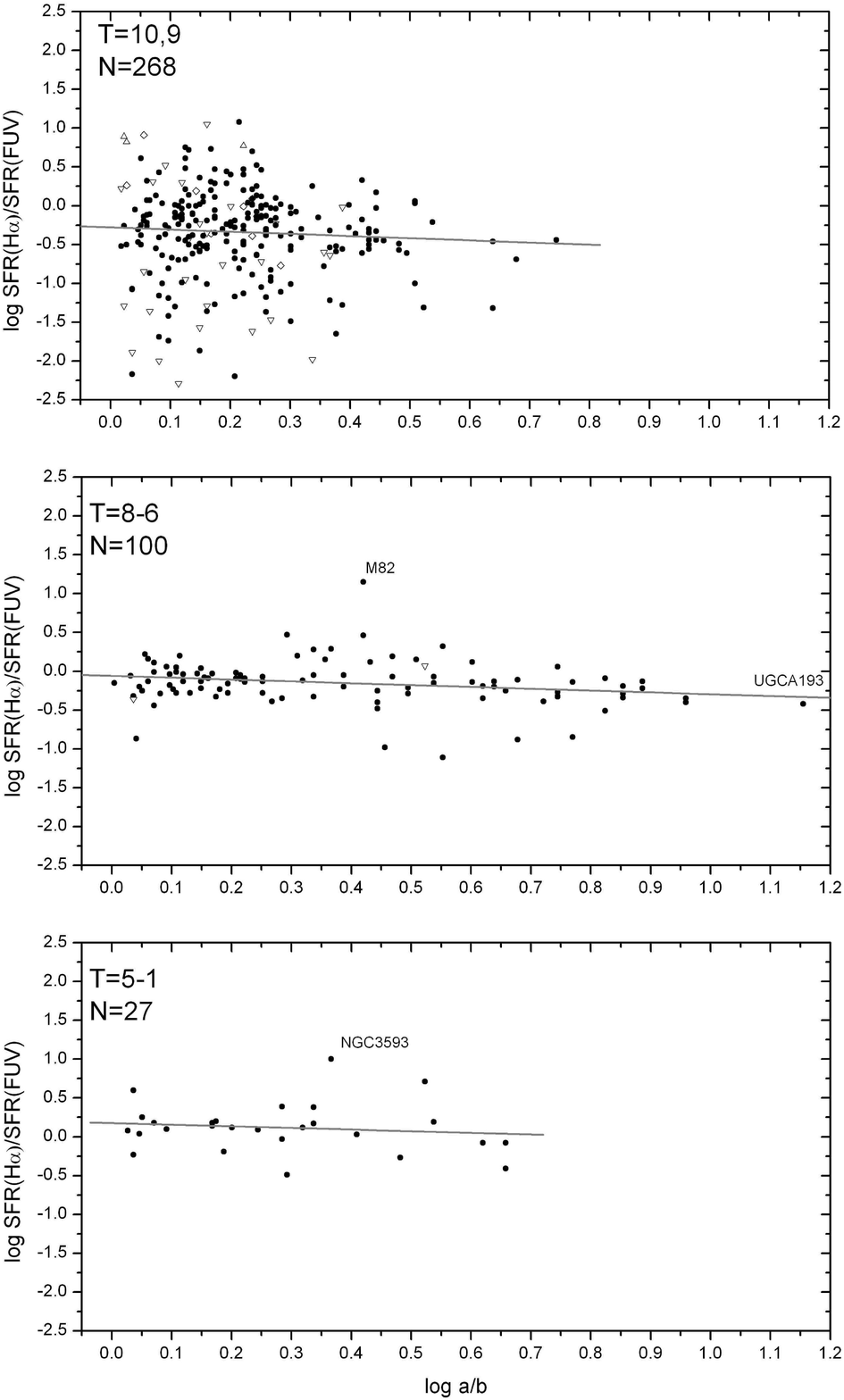} 
\caption{SFR (\Halpha{})/SFR(FUV) versus the apparent axial ratio for the LV galaxies of dIr, dIm, and BCD types (the top panel), late-type spirals (the middle panel), and early-type spirals (the bottom panel).} 
\end{figure*} 

\begin{figure*} 
\includegraphics[scale=1.0]{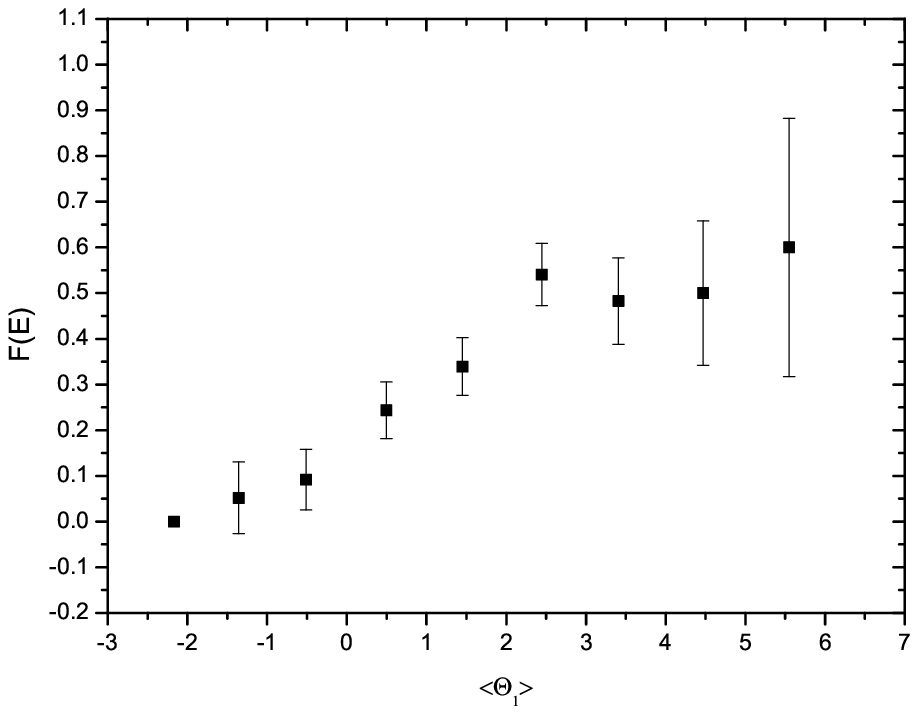}\\ \includegraphics[scale=1.0]{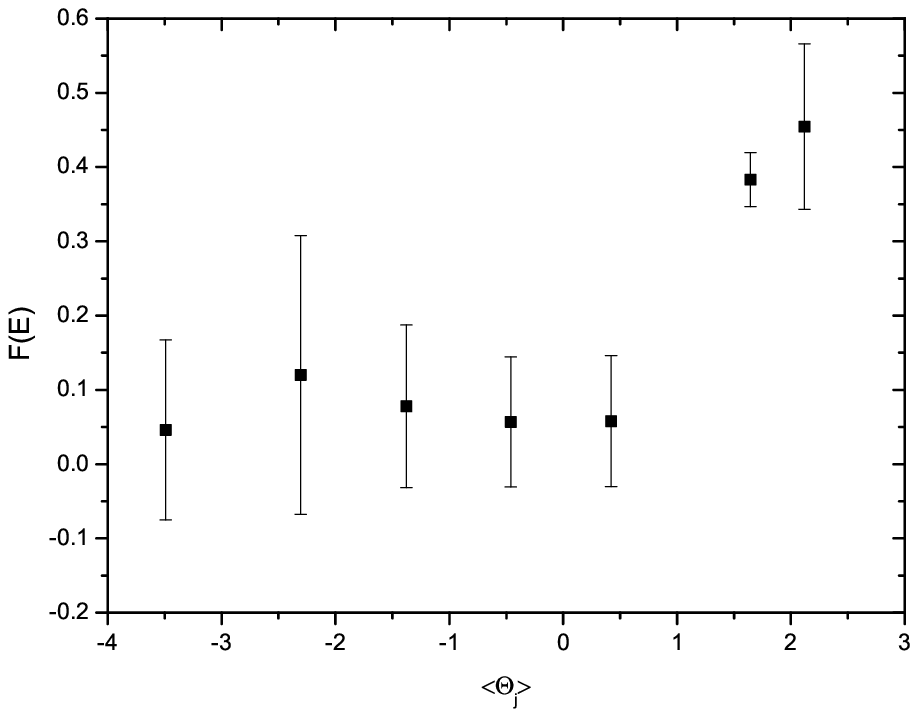} 
\caption{Fraction of (E, S0, dSph) galaxies versus the tidal index $\Theta_1$ (the top panel) and the local density contrast $\Theta_j$ (the bottom panel). Vertical bars indicate standard errors.} 
\end{figure*} 

\begin{figure*} 
\includegraphics[scale=1.0]{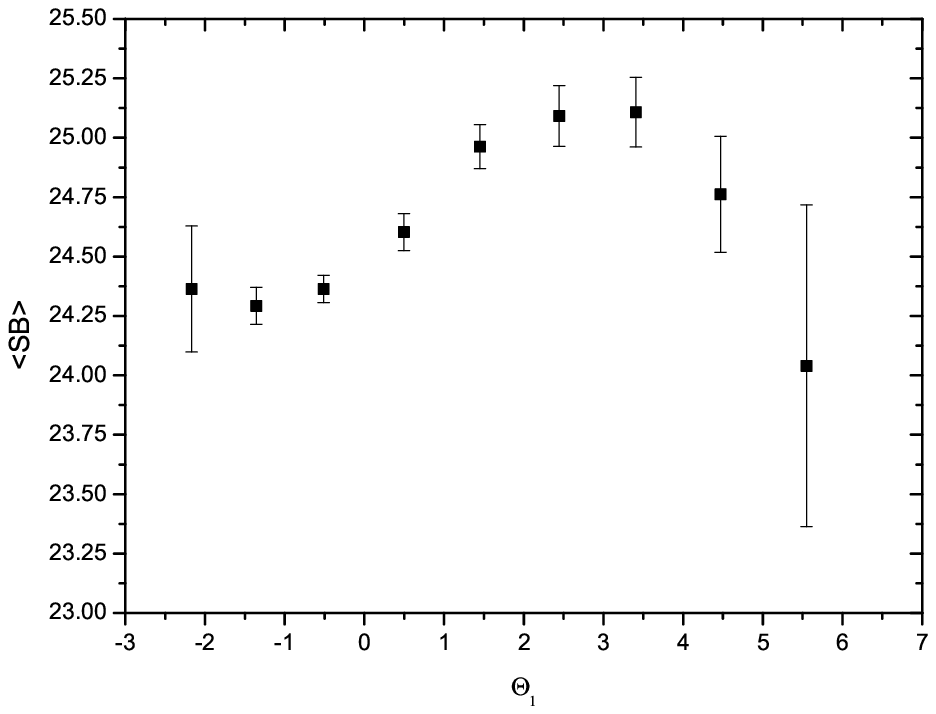}\\ \includegraphics[scale=1.0]{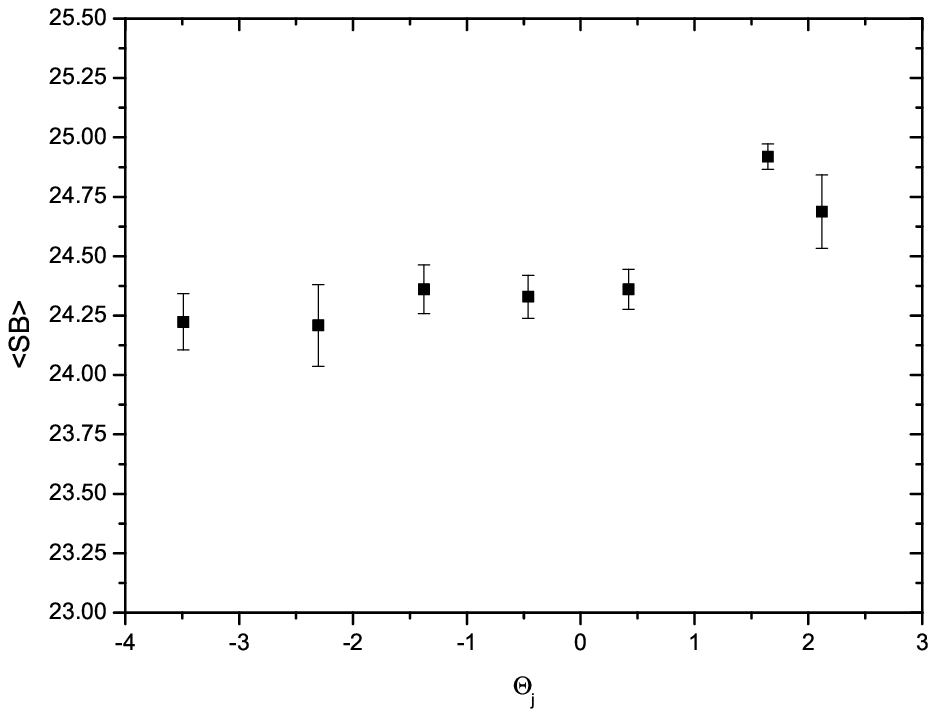} 
\caption{Mean surface brightness in the B band versus $\Theta_1$ (the upper panel) and $\Theta_j$ (the lower panel). Vertical bars indicate standard errors.} 
\end{figure*} 

\appendix
\section{List of 1029 galaxies in the Local Volume with distance estimates
D < 11.0 Mpc and the Galactic B-band extinction $A_B$ < 3.0 mag.}
\begin{tabular}{lcrrrrrrrrr}
\hline\hline
 Name& RA (J2000.0) Dec. &T &D &log M* &log($M_{\rm HI}$)&log(SFR[FUV]) &log(SFR[\Halpha{}]) &SB &$\Theta_1$ &$\Theta_j$\\ 
 \hline
UGC12894   &000022.5$+$392944 &$10$ &8.47 &7.58 & 7.92   &$ -2.03$   &$-2.29$  &25.2 &$-1.32$ &$ 0.12$\\
WLM        &000158.1$-$152740 &$ 9$ &0.98 &7.70 & 7.84   &$ -2.23$   &$-2.68$  &24.8 &$-0.04$ &$ 1.75$\\
And XVIII  &000214.5$+$450520 &$-3$ &1.31 &6.44 &$<$6.65 &$<-5.84$   &$<-5.97$&26.8 &$ 0.54$ &$ 1.54$\\
PAndAS-03  &000356.4$+$405319 &$-3$ &0.78 &4.38 &        &$<-6.35$   &         &27.8 &$ 2.78$ &$ 1.75$\\
PAndAS-04  &000442.9$+$472142 &$-3$ &0.78 &5.59 &        &$ -5.67$   &$<-6.82$ &23.1 &$ 2.49$ &$ 1.75$\\
PAndAS-05  &000524.1$+$435535 &$-3$ &0.78 &4.75 &        &$<-6.38$   &         &25.6 &$ 2.77$ &$ 1.75$\\
ESO409-015 &000531.8$-$280553 &$ 9$ &8.71 &8.10 &8.10    &$ -1.72$   &$-1.47$  &24.1 &$-1.98$ &$-2.05$\\
AGC748778  &000634.4$+$153039 &$10$ &6.22 &6.39 &6.64    &$ -3.53$   &$-4.64$  &24.9 &$-1.87$ &$-2.72$\\
And XX     &000730.7$+$350756 &$-3$ &0.80 &5.26 &        &$ -5.96$   &$<-6.34$ &27.0 &$ 2.43$ &$ 1.75$\\
UGC00064   &000744.0$+$405232 &$10$ &9.60 &8.15 &8.58    &$ -1.63$   &$-1.51$  &25.0 &$-1.88$ &$-0.96$\\
ESO349-031 &000813.3$-$343442 &$10$ &3.21 &7.12 &7.13    &$ -3.02$   &$-4.03$  &24.7 &$ 0.15$ &$ 1.75$\\
NGC0024    &000956.4$-$245748 &$ 5$ &7.31 &9.48 &8.64    &$ -0.60$   &$-1.01$  &24.5 &$-1.20$ &$ 0.07$\\
\hline
\multicolumn{11}{l} {Notes. Morphological type in de Vaucouleurs scale, distance $D$ in $Mpc$, stellar mass $M*$}\\ 
\multicolumn{11}{l} {and  hydrogen mass $M_{\rm HI}$ in Solar mass units, SFR from FUV-flux and \Halpha{} flux in ($M_{sun}/yr$), }\\
\multicolumn{11}{l} {average B-band surface brightness in $mag/arcsec^2$, $\Theta_1$ and $\Theta_j$ are dimensionless }\\
\multicolumn{11}{l} {parameters characterizing the local density contrast.}\\ 
\end{tabular} 

\end{document}